# Anomalously high supercurrent density in a two-dimensional topological material

Qi Zhang[1*†], Md Shafayat Hossain[1*†], Brian Casas[2], Wenkai Zheng[2], Zi-Jia Cheng[1], Zhuangchai Lai[3,4], Yi-Hsin Tu[5], Guoqing Chang[6], Yao Yao[3], Siyuan Li[3], Yu-Xiao Jiang[1], Sougata Mardanya[5], Tay-Rong Chang[5], Jing-Yang You[7], Yuan-Ping Feng[7,8], Guangming Cheng[9], Jia-Xin Yin[1], Nana Shumiya[1], Tyler A. Cochran[1], Xian P. Yang[1], Maksim Litskevich[1], Nan Yao[9], Kenji Watanabe[10], Takashi Taniguchi[11], Hua Zhang[3,12,13†], Luis Balicas[2], M. Zahid Hasan[1,14†]

[1]*Laboratory for Topological Quantum Matter and Advanced Spectroscopy (B7), Department of Physics, Princeton University, Princeton, New Jersey, USA.*
[2]*National High Magnetic Field Laboratory, Tallahassee, Florida 32310, USA.*
[3]*Department of Chemistry, City University of Hong Kong, Hong Kong, China.*
[4]*Department of Applied Physics, The Hong Kong Polytechnic University, Hong Kong, China.*
[5]*Department of Physics, National Cheng Kung University, 701 Tainan, Taiwan*
[6]*Division of Physics and Applied Physics, School of Physical and Mathematical Sciences, Nanyang Technological University, Singapore 637371, Singapore.*
[7]*Department of Physics, National University of Singapore, 2 Science Drive 3, Singapore 117551, Singapore*
[8]*Centre for Advanced 2D Materials, National University of Singapore, 6 Science Drive 2, Singapore 117546, Singapore*
[9]*Princeton Institute for Science and Technology of Materials, Princeton University, Princeton, NJ, USA.*
[10]*Research Center for Functional Materials, National Institute for Materials Science, Tsukuba, Japan.*
[11]*International Center for Materials Nanoarchitectonics, National Institute for Materials Science, Tsukuba, Japan.*
[12]*Hong Kong Branch of National Precious Metals Material Engineering Research Center (NPMM), City University of Hong Kong, Hong Kong, China.*
[13]*Shenzhen Research Institute, City University of Hong Kong, Shenzhen 518057, China.*
[14]*Lawrence Berkeley National Laboratory, Berkeley, California 94720, USA.*

*\* These authors contributed equally to this work*

*†Corresponding to: qz9@princeton.edu; mdsh@princeton.edu; hua.zhang@cityu.edu.hk; mzhasan@princeton.edu*

Ongoing advances in superconductors continue to revolutionize technology thanks to the increasingly versatile and robust availability of lossless supercurrent. In particular high supercurrent density can lead to more efficient and compact power transmission lines, high-field magnets, as well as high-performance nanoscale radiation detectors and superconducting spintronics. Here, we report the discovery of an unprecedentedly high superconducting critical current density (17 MA/cm$^2$ at 0 T and 7 MA/cm$^2$ at 8 T) in 1T′-WS$_2$, exceeding those of all reported two-dimensional superconductors to date. 1T′-WS$_2$ features a strongly anisotropic (both in- and out-of-plane) superconducting state that violates the Pauli paramagnetic limit signaling the presence of unconventional superconductivity. Spectroscopic imaging of the vortices further substantiates the anisotropic nature of the superconducting state. More intriguingly, the normal state of 1T′-WS$_2$ carries topological properties. The band structure obtained via angle-resolved photoemission spectroscopy and first-principles calculations points to a Z$_2$ topological invariant. The concomitance of topology and superconductivity in 1T′-WS$_2$ establishes it as a topological superconductor candidate, which is promising for the development of quantum computing technology.

Since the discovery of superconductivity by Heike Kamerlingh Onnes back in 1911 [1], superconductors have revolutionized science and technology through numerous applications ranging from superconducting qubits to high-field magnets [2-4]. High-field magnets fabricated from superconductors with high critical current density, have enabled scientific discoveries across physical, chemical, and biological sciences [5-7]. On the other hand, superconducting materials exhibiting topological properties offer possibilities beyond this classical application paradigm, opening a new frontier to implement fault-tolerant quantum information technologies. Recently, two-dimensional (2D) transition metal dichalcogenides (TMDCs) attracted considerable interests thanks to their abundant crystal structures and novel physical properties [8-11]. Specifically, hole-doped TMDCs have been considered as candidates for topological superconductivity based on momentum-space-split spinless fermions [8]. For example, the coexistence of superconductivity with a topologically non-trivial electronic state makes 2M-WS$_2$ a good candidate for topological superconductivity [12]. Here, we access both avenues and demonstrate an unprecedentedly high superconducting critical current density and topological features in the 2D superconductor 1T′-WS$_2$.

1T′-WS$_2$ is composed of a distorted [WS$_6$] octahedral and crystallizes in a monoclinic layered structure [13], as shown in Fig. 1a. High purity 1T′-WS$_2$ crystals were synthesized via a previously reported method [13]. The single-phase nature can be observed in the cross-sectional scanning transmission electron microscope (STEM) image (Fig. 1b). STEM image unveils the atomic stacking pertaining to a monoclinic and distorted structure. This atomic-resolution characterization confirms the high crystallinity and phase purity of the as-synthesized 1T′-WS$_2$ crystals, consistent with the previous report [13]. After characterizing the bulk material, we fabricated devices based on few-layer 1T′-WS$_2$ for transport measurements. Thin flakes of 1T′-WS$_2$ obtained via mechanical exfoliation were transferred onto a SiO$_2$ (285 nm)/Si substrate (inset of Fig. S1a). The Raman spectrum of the as-prepared 1T′-WS$_2$ flake shows a series of peaks at ~112, ~178, ~270, ~316, and ~407 cm$^{-1}$ (Fig. S1a), consistent with single-phase 1T′-WS$_2$ [13]. The thickness of the flakes used in our measurements is ~6.1 nm as measured by the atomic force microscope (Fig. S1b). The device was fabricated following a Hall-bar configuration and measured from $T$ = 300 K to 2.0 K in a Physical Property Measurement System. Figure 1c depicts the four-probe resistance as a function of temperature and captures the electrical transport behavior of the sample. At high temperatures, it exhibits metallic behavior (d$R$/d$T$ > 0), indicating phonon-scattering-dominated transport [14]. The superconducting transition occurs at 7.7 K, which is slightly lower than the bulk critical temperature ($T_c$) of 1T′-WS$_2$ (8.6 K) [13]. We also measured the Hall effect of 1T′-WS$_2$ above the critical temperature. Strikingly, the carrier concentration in 1T′-WS$_2$ approaches $10^{15}$~$10^{16}$ cm$^{-2}$ at $T$ = 10 K (Fig. S2a and S2b). This value is much higher than the typical carrier concentration (~$10^{14}$ cm$^{-2}$) of 2D superconductors with electrostatic gating [15].

To investigate the superconducting state of 1T′-WS$_2$, we performed magneto-transport measurements (Fig. 1d). We start with the angular dependence of the upper critical magnetic field ($H_{c2}$), defined as the magnetic field at which the resistance drops to 50% of its normal state value. The details of the angular dependent measurement are described in Fig. S3. For a clear visualization, we normalized the resistance by $R_n$, i.e., the normal state resistance for all the samples. Figure 1e summarizes the magnetic field dependence of the resistance at different angles ($\theta$) at $T$ = 0.33 K, where $\theta$ is the angle between the $z$-axis and the magnetic field direction (Fig. 1d). As the sample is rotated from the perpendicular ($\theta$ = 0°) to a parallel field ($\theta$ = 90°) configuration, the transition towards superconductivity progressively shifts to higher fields, manifesting a clear superconducting anisotropy (Fig. 1e). In Fig. 1f, we present a plot of $H_{c2}$ as a function of $\theta$, showing that the highest $H_{c2}$ occurs when the magnetic field is applied parallel to the sample plane. To understand the anisotropic nature of $H_{c2}$, we fitted our data to the Tinkham formula, which describes the angular dependence of $H_{c2}$ for a 2D superconductor [16]:

$$\left|\frac{H_{c2}(\theta)\cos\theta}{H_{c2,\perp}}\right| + \left(\frac{H_{c2}(\theta)\sin\theta}{H_{c2,//}}\right)^2 = 1$$

where $H_{c2,\perp}$ and $H_{c2,//}$ are the upper critical field for fields perpendicular and parallel to the plane of the sample, respectively. As shown in Fig. 1f, the blue fitting curve matches the data quite well and thus confirms the 2D nature of the superconductivity in 1T′-WS$_2$. The fitting of the angle dependent critical field for smaller angular regimes to the 2D Tinkham formula is shown in Fig. S4.

After exploring the anisotropy of $H_{c2}$ along the out-of-plane directions, we then examined how $H_{c2}$ evolved along the in-plane directions. As the device is rotated from the $x$-axis ($\varphi$ = 0°; $\varphi$ is the angle between the magnetic field and the $x$-axis as shown in Fig. 1d) to the $y$-axis ($\varphi$ = 90°), the superconducting transition progressively shifts from higher fields to lower fields (Fig. 1g). Careful measurements were performed to rule out the possibility of an accidental out-of-plane component (Fig. S5). Such a planar anisotropy is likely to result from the reduced crystal symmetry due to the distorted structure of 1T′-WS$_2$, as clearly seen in Fig. 1a. Figure 1h, which shows $H_{c2}$ as a function of $\varphi$, reveals an emergent two-fold symmetry. Furthermore, we observed that the largest value of the in-plane $H_{c2}$ (28 T). To obtain a quantitative understanding of such a large value, we compared it to the expected Pauli paramagnetic limiting field. In conventional superconductors, a sufficiently high external magnetic field can suppress superconductivity through the orbital [17] and spin Zeeman effect [18,19]. For a few-layer sample, the suppression from the orbital effect is nearly absent when the magnetic field is parallel to the sample plane. Consequently, the Zeeman effect imposes an upper bound on $H_{c2}$, known as the Pauli limit ($H_p$ = 1.84×$T_c$ T/K) [20]. We find that the in-plane $H_{c2}$ (28 T) in 1T′-WS$_2$ clearly violates the Pauli limit (14 T for $T_c$ = 7.7 K). Such a violation combined with the emergence of two-fold symmetry for the in-plane $H_{c2}$ suggests unconventional superconductivity in 1T′-WS$_2$.

We further explored the superconducting transition via systematic temperature dependent measurements. Figures 1i and 1j show such data taken when the magnetic field was perpendicular and parallel to the sample plane, respectively. In both cases, the superconducting transition shifts gradually to lower magnetic fields as the temperature increases. The temperature dependence of the out-of-plane upper critical field ($H_{c2,\perp}$) and in-plane upper critical field ($H_{c2,//}$) are summarized in Fig. 1k. $H_{c2,\perp}$ displays a linear dependence on temperature, that is well fitted by the standard Ginzburg-Landau (GL) theory for 2D superconductors [16]:

$$H_{c2,\perp}(T) = \frac{\Phi_0}{2\pi\xi_{GL}(0)^2}\left(1 - \frac{T}{T_c}\right),$$

where $\xi_{GL}(0)$ is the zero-temperature GL in-plane coherence length, $\Phi_0$ is the magnetic flux quantum, and $T_c$ is the critical temperature at which the resistance drops to 50% of its value in the normal state. From the fit we can estimate the coherence length $\xi_{GL}(0) \approx 9.6$ nm. The temperature dependence of $H_{c2,//}$, on the other hand, follows the GL formula expected for 2D superconductors [16]:

$$H_{c2,//}(T) = \frac{\Phi_0\sqrt{3}}{\pi\xi_{GL}(0)d_{sc}}\left(1 - \frac{T}{T_c}\right)^{1/2},$$

where $d_{SC}$ is the superconducting thickness. From the fitting of $H_{c2,//}$, the superconducting thickness is around 3.2 nm, which is smaller than $\xi_{GL}(0)$ and consistent with 2D superconductivity.

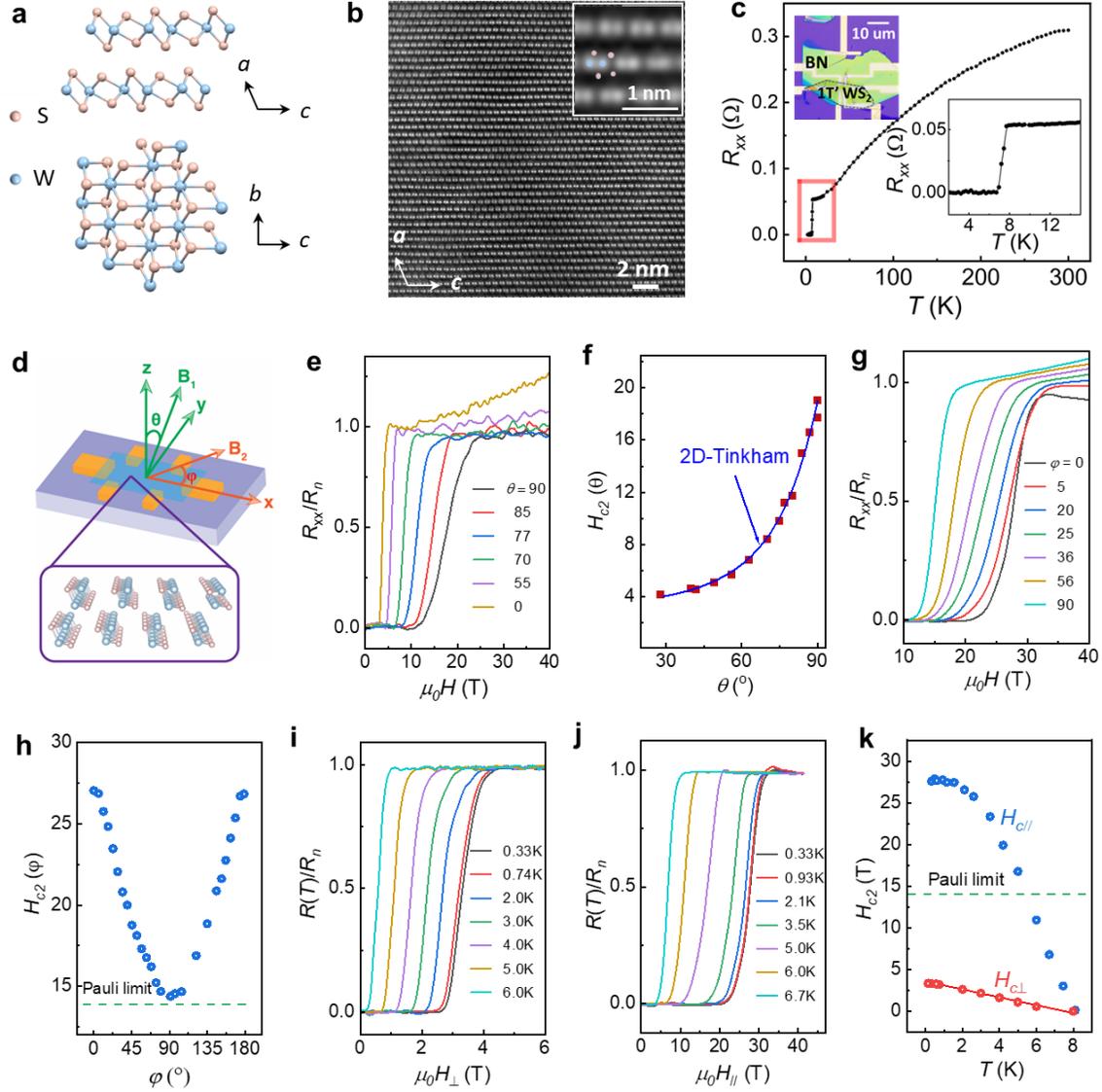

**FIG.1 Crystal structure and Superconductivity of 1T′-WS$_2$. a,** Schematic illustration of the structure of 1T′-WS$_2$. Top panel: side view of the crystallographic structure; bottom panel: top view of a typical monolayer. **b,** Cross-sectional STEM image of 1T′-WS$_2$. Inset: high-magnification STEM image of layered structure with atomic resolution. **c,** Temperature-dependent electrical resistance of the mechanically exfoliated 1T′-WS$_2$ without magnetic field. Insets: optical image of the 1T′-WS$_2$ device covered by h-BN with Hall-bar configuration (top) and small range $R_{xx}$-$T$ plot of 1T′-WS$_2$ around $T_c$ shown in the area within the red rectangle (bottom). **d,** Schematic illustration of a 1T′-WS$_2$ device and the rotation experiment setup, where the $x$-axis is parallel to $c$-axis of the crystal and $z$-axis is perpendicular to crystalline plane. $\theta$ is the angle between the out-of-plane magnetic field and the $z$-axis; $\varphi$ is the angle between the in-plane magnetic field and the $x$-axis. **e,** Magnetic field dependence of the normalized resistance of the 1T′-WS$_2$ device at $T$ = 0.33 K with different out-of-plane rotation angles $\theta$. **f,** The $\theta$-dependence of the upper critical field. The blue curve denotes a fit to the data following the Tinkham formula for a 2D superconductor. **g,** Magnetic field dependence of the normalized resistance of the 1T′-WS$_2$ device at $T$ = 0.33 K with different in-plane rotation angles $\varphi$. **h,** The $\varphi$-dependence of the upper critical field. The green dashed line indicates the Pauli limit. **i, j,** Superconducting transition of the 1T′-WS$_2$ device under a perpendicular magnetic field (**i**) and under a parallel magnetic field (**j**) at different temperature. **k,** Temperature dependence of the upper critical field with magnetic field directions parallel and perpendicular to the crystal plane. The red curve represents the linear relationship between $H_{c2,\perp}$ and $T$ according to the 2D GL theory.

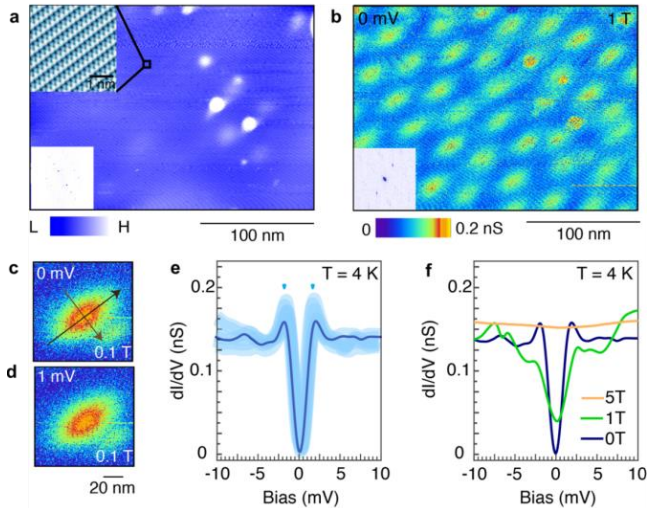

**FIG.2 Scanning tunneling microscopy measurements on 1T′-WS₂. a**, Topographic image of the *bc* plane of 1T′-WS₂. Top inset: a zoom-in view of the topographic image showing the atomic arrangements. Bottom inset: Fast Fourier transform of the topographic image. **b**, A zero-bias conductance map of vortices at 1 T. Inset: Fourier transform of the d*I*/d*V* map. **c, d**, The conductance map of a single vortex at 1 T with zero bias (**c**) and 1 mV bias (**d**). **e**, Tunneling spectroscopy spectrum taken at 4.2 K, revealing a superconducting gap. Light blue curves are the differential spectra taken at different positions on the surface; the dark blue curve denotes the average spectra. **f**, Field dependence of tunneling spectroscopy taken at 0 T, 1 T and 5 T.

To further characterize the anisotropic superconductivity in 1T′-WS$_2$, we performed scanning tunneling microscopy (STM) measurements and directly imaged the vortices under magnetic field. A single crystal was cleaved in-situ at $T = 77$ K and measured at $T = 4.2$ K. Figure 2a shows the topography of 1T′-WS$_2$ over a large area. The atomically resolved STM topographic image reveals a clean surface featuring zigzag chains along the *b*-axis of the crystal (top inset of Fig. 2a). In addition, the corresponding fast Fourier transform pattern also exhibits the distorted octahedral coordination feature (bottom inset of Fig. 2a). A zero-energy conductance map under 1 T applied perpendicularly to the *bc* plane is shown in Fig. 2b. The Fourier transform of the d*I*/d*V* map is two-fold symmetric (inset of Fig. 2b). The conductance maps of a single vortex at 0.1 T taken at *V* = 0 mV (Fig. 2c) and 1 mV (Fig. 2d) further highlight the anisotropic nature of the superconductivity. Consistent with the anisotropy observed in our transport data, the vortices are anisotropic and elongated along the *b* direction, reflecting the anisotropy of the Ginzburg-Landau coherence length between both directions. Tunneling differential conductance collected from an atomically resolved lattice illustrates a superconducting gap with sharp coherence peaks (Fig. 2e). This superconducting gap disappears gradually as the magnetic field is increased (Fig. 2f).

Subsequently, we performed critical current density ($J_c$) measurements. As alluded in the introduction, an important aspect of a superconductor is its $J_c$, which dictates several practical applications. The higher the $J_c$ of a superconductor, the smaller and more efficient the superconducting devices that can be fabricated from it or the larger the magnetic fields that can be generated. We measured differential resistance of the 1T′-WS$_2$ device with thickness of 6 nm as a function of direct current (DC) bias current at different temperatures (Fig. 3a). Note that, $J_c$ is defined as the current density at which the differential resistance (d*V*/d*I*) reaches its maximum, as reported in previous works [21,22]. Remarkably, as seen in Fig. 3b, 1T′-WS$_2$ exhibits ultrahigh critical current densities reaching 17 MA/cm$^2$ at $T = 0.33$ K. Figure 3b highlights the temperature dependence of the critical current density, featuring an enormous $J_c = 13$ MA/cm$^2$ at liquid He temperature (4.2 K). In addition, we systematically measured the critical currents of samples with different layer thicknesses, as shown in Fig. S6. The thickness dependence of the critical current density is summarized in Fig. 3c. There is no obvious difference among the samples with thicknesses exceeding 20 nm. The critical current densities increase as the devices become thinner, which is also observed in atomically thin TaS$_2$ [23]. Furthermore, we evaluated the field dependence of $J_c$ (Fig. S7). The critical current density falls rapidly as the perpendicular magnetic field increases (Fig. 3d). In contrast, the critical current density is rarely influenced by a parallel magnetic field since 1T′-WS$_2$ shows extremely high in-plane upper critical fields. Even under an 8 T in-plane magnetic field, $J_c$ is substantially large (7 MA/cm$^2$).

Experimentally, numerous 2D superconducting transition metal dichalcogenides have been studied [20-33]. In-plane anisotropic upper critical fields were observed in 2H-NbSe$_2$ [24] and T$_d$-MoTe$_2$ [25]. 2H-NbSe$_2$ [20] and ionic-gated 2H-MoS$_2$ [26] also exhibited high in-plane upper critical fields. However, we emphasize that 1T′-WS$_2$ is the only 2D material to our knowledge that shows the suitable critical temperature and high critical current under high in-plane magnetic field, which are crucial for building high-field magnets. Even for a thick sample, the in-plane critical field surpasses 8 T at 4 K (Fig. S8). We summarize the parameters of 2D superconductors in Fig. 3e. As for 1T-MoS$_2$ [27], 2H-TaS$_2$ [23], 3R-TaSe$_2$ [28], T$_d$-MoTe$_2$ [25], 2H-NbS$_2$ [29], their critical temperatures are below the temperature of liquid helium (4.2 K), rendering the construction of high-field magnets impractical. Gated MoS$_2$ displays a relatively high critical temperature and also very high critical fields, but superconductors under ionic gating are not suitable for applications [30]. Lastly, 2H-NbSe$_2$ is comparable to 1T′-WS$_2$ in critical fields and critical temperatures. However, its critical current density is two orders of magnitude lower than that of 1T′-WS$_2$ [31]. The significance of our work is that we report an unprecedentedly high superconducting critical current density (17 MA/cm$^2$ at 0 T) in 1T′-WS$_2$, which exceeds those of all the known 2D superconductors to date [21-33]. Notably, it even exceeds the $J_c$ of MgB$_2$ films [34], a well-known superconductor for high-critical-current applications (Fig. 3e). Even under an 8 T in-plane magnetic field, the $J_c$ of 1T′-WS$_2$ is substantially large (7 MA/cm$^2$). As a reference, the critical currents of commercial magnet building materials are listed here, such as Nb-Ti alloy (0.1 MA/cm$^2$ at 10 T) and Nb$_3$Sn (0.5 MA/cm$^2$ at 10 T) [35]. The large $J_c$ at zero and finite magnetic fields makes 1T′-WS$_2$ a potential candidate for future study on building next-generation superconducting magnets.

Having explored the superconductivity of 1T′-WS$_2$, we turn to the topological features pertinent to its electronic band structure using a series of theoretical calculations and angle-resolved photoemission spectroscopy (ARPES) experiments. The calculated bulk band structure is shown in Fig. S9. Besides the continuous energy gap between conduction band and valence band around the Fermi level, we observe a band inversion at the Γ-point between W $d$ and S $p$ orbitals, which leads to a strong topological insulating phase. Furthermore, the surface-projected calculation shows the topological Dirac surface state emerging from the valence band and merging into conduction bands (Fig. 4a). The corresponding ARPES data (Fig. 4b), taken at $T = 10$ K (above $T_c$) matches the first principles calculations below $E_F$. In particular we identify the linear-dispersed hole pocket at Γ to be the lower cone of the topological surface state, as it shows no photon-energy dependence and agrees well with the calculated dispersion of the Dirac state (More details of ARPES data analysis are shown in Figs. S10-12). ARPES Fermi surface map also visualizes the highly anisotropic Fermi surface (Fig. 4c), which possibly contributes to the extremely anisotropic $H_{c2}$ in 1T′-WS$_2$. The calculated superconducting gap of 1T′-WS$_2$ on the Fermi surface is presented in Fig. 4d. These results lend crucial credence to the in-plane anisotropy of superconductivity. It is noted that so far there is no clear relationship between the topological nature and high critical current.

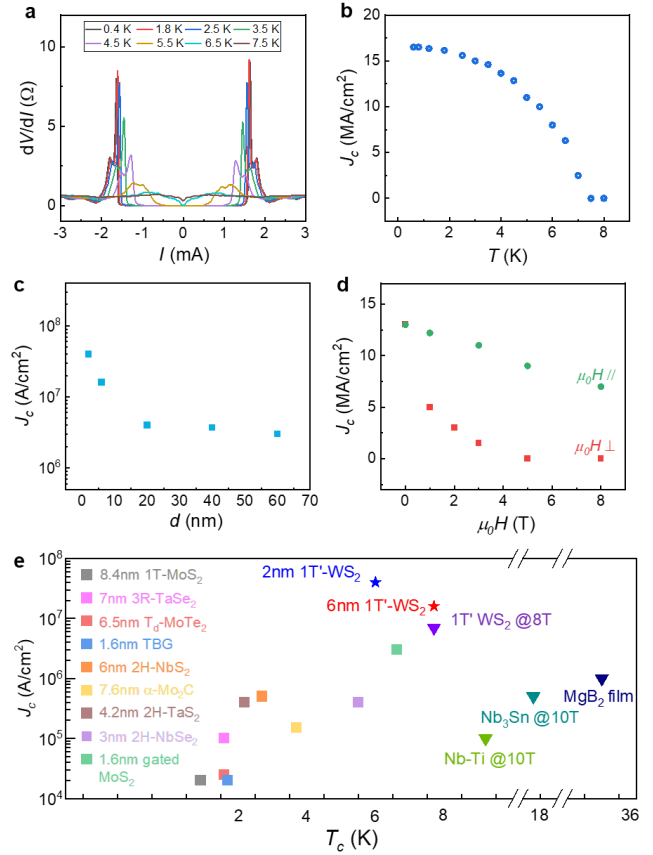

**FIG.3 Ultrahigh critical current density of 1T′-WS$_2$. a**, Differential resistance of a 6-nm-thick 1T′-WS$_2$ sample as a function of the direct current (DC) bias at different temperatures. **b**, Critical current density for a 1T′-WS$_2$ device as a function of the temperature. **c**, Critical current density of the 1T′-WS$_2$ device plotted as a function of sample thickness. **d**, Critical current densities of the 1T′-WS$_2$ device plotted as a function of perpendicular and parallel magnetic fields. **e**, Comparison of critical current densities among 1T′-WS$_2$ and other representative 2D superconductors, such as twisted bilayer graphene (TBG) and transition metal dichalcogenides. Commercial magnet building materials are also included for reference. Here, the superconducting critical temperatures $T_c$ of the different materials were determined under zero magnetic field.

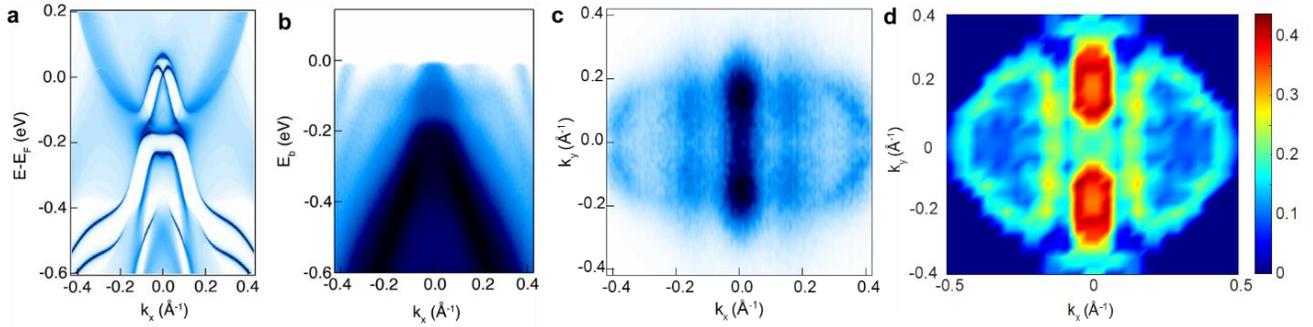

**FIG.4 Topological features of 1T′-WS$_2$. a**, Calculated surface band structure of 1T′-WS$_2$ at $k_y = 0$, featuring a topological Dirac surface state near the Fermi level. **b**, Energy-momentum cut acquired through ARPES. **c**, Fermi surface of 1T′-WS$_2$. **d**, Calculated superconducting gap which all $k_z$ are projected in the surface Brillouin zone at 2.5 K on the Fermi surface. The unit of the color bar is meV.

In summary, combining a series of experimental and numerical techniques, we comprehensively studied 1T′-WS$_2$ and find a unique blend of ultrahigh critical supercurrent density, large superconducting anisotropy (in-plane versus out-of-plane) along with topological features. Our findings not only provide a promising material platform for high magnetic field technologies but also unveil a promising platform for future exploration of topological superconductivity, which may be used to fabricate topologically protected qubits for future quantum computing schemes.

ACKNOWLEDGEMENTS. Experimental and theoretical work at Princeton University was supported by the Gordon and 286 Betty Moore Foundation (GBMF4547; M.Z.H.). The material characterization is supported by the United States 287 Department of Energy (US DOE) under the Basic Energy Sciences program (grant number DOE/BES DE-FG-288 02-05ER46200). L.B. is supported by DOE-BES through award DE-SC0002613. The National High Magnetic Field Laboratory acknowledges support from the US-NSF Cooperative agreement Grant number DMR-1644779 and the state of Florida. The authors acknowledge the sample characterization of Imaging and Analysis Center (IAC) at Princeton University, partially supported by the Princeton Center for Complex Materials (PCCM) and the NSF-MRSEC program (MRSEC; DMR-2011750). G.C. acknowledges the support of the National Research Foundation, Singapore under its Fellowship Award (NRF-NRFF13-2021-0010) and the Nanyang Assistant Professorship grant from Nanyang Technological University. J.Y.Y. and Y.P.F. is supported by the Ministry of Education, Singapore, under its MOE AcRF Tier 3 Award MOE2018-T3-1-002. H.Z. thanks the support from ITC via the Hong Kong Branch of National Precious Metals Material Engineering Research Center (NPMM), the Research Grants Council of Hong Kong (AoE/P-701/20), the Start-Up Grant (Project No. 9380100) and grant (Project No. 1886921) from the City University of Hong Kong, and the Science Technology and Innovation Committee of Shenzhen Municipality (grant no. JCYJ20200109143412311). K.W. and T.T. acknowledge support from JSPS KAKENHI (Grant Numbers 19H05790, 20H00354 and 21H05233).